# Influence of Phonon Scattering on the Performance of p-i-n Band-to-Band-Tunneling Transistors

Siyuranga O. Koswatta[*] and Mark S. Lundstrom

School of Electrical and Computer Engineering, Purdue University, West Lafayette, IN

47907-1285

Dmitri E. Nikonov

Technology and Manufacturing Group, Intel Corp., SC1-05, Santa Clara, CA 95052

*Abstract* – Power dissipation has become a major obstacle in performance scaling of modern integrated circuits, and has spurred the search for devices operating at lower voltage swing. In this letter, we study p-i-n band-to-band tunneling field effect transistors (TFET) taking semiconducting carbon nanotubes as the channel material. The on-current of these devices is mainly limited by the tunneling barrier properties, and phonon scattering has only a moderate effect. We show, however, that the off-current is limited by phonon absorption assisted tunneling, and thus is strongly temperature-dependent. Subthreshold swings below the 60mV/decade conventional limit can be readily achieved even at room temperature. Interestingly, although subthreshold swing degrades due to the effects of phonon scattering, it remains low under practical biasing conditions.

---

[*] Email: koswatta@purdue.edu

As integrated circuit densities continue to increase, power dissipation has become a critical issue for chip designers. Silicon metal-oxide-semiconductor field-effect transistors (MOSFETs) now operate with a power supply voltage of approximately 1V [1]. A key challenge for device research is to develop high performance transistors that operate at substantially lower voltages. To maintain high on-currents at lower voltages, it is likely that devices will need to operate with the subthreshold swing below the conventional MOSFET limit of 60mV/decade (at room temperature). Band-to-band tunneling (BTBT) transport in devices has been proposed [2-7] and demonstrated [8-13] as one means to produce low-voltage transistors. The BTBT FET device concept is currently being explored in a number of different materials and structure options including that based on carbon nanotubes (CNTs) [14-19]. CNTs are especially promising for such devices because their small effective masses and direct bandgap [20] promote BTBT. Furthermore, physically-detailed simulations that include a full treatment of quantum transport in the presence of dissipative phonon scattering have been developed [21, 22] and thoroughly benchmarked against experimental results [23-25], so a tool to explore device design and optimization is available.

Our objective in this letter is to identify the physical mechanisms that limit the performance of a BTBT FET; specifically, the maximum on-state current, off-state current, and the steepness of the on-off transition. We use the carbon nanotube BTBT FET as a model device to address these questions because of the aforementioned benefits of this system. Our earlier work has shown that BTBT in a CNT-MOSFET type structure is dominated by phonon assisted inelastic tunneling that leads to degradation of the desirable device characteristics [24, 26]. Here, we examine another popular tunneling device structure based on the gated p-i-n geometry (p-i-n TFET) [2-5, 7, 15, 16]. The purpose of this letter is to examine how phonon scattering affects the performance of such devices. Design optimization and the comparison of performance against conventional MOSFETs will be the subject of a future publication. Although we use the CNT as a model device, the general conclusions are expected to be broadly applicable to this class of devices fabricated in different materials.

The model device structure used in this study, shown in figure 1, has a cylindrical wrap-around gate and doped source/drain regions (see caption for device parameters). We have performed dissipative quantum transport calculations using the non-equilibrium Green's function (NEGF) formalism [27] along with self-consistent electrostatics. A detailed description of the simulation procedure is presented in [22], and here we summarize the main equations for the sake of clarity. The device Green's function, $G$, at an energy $E$ in the presence of electron-phonon (e-ph) scattering is given by [27],

$$G = \left[ EI - H_{pz} - \Sigma_S - \Sigma_D - \Sigma_{scat} \right]^{-1} \quad (1)$$

where, $I$ is the identity matrix, and $H_{pz}$ is the device Hamiltonian in the nearest-neighbor tight-binding description [20]. Here, the mode-space treatment for carrier transport is used [22, 28], in that, we are considering the lowest conduction band and the highest valence band with two-fold spin and two-fold valley degeneracies [20]. $\Sigma_{S/D}$ and $\Sigma_{scat}$ are the self-energy functions (energy dependent) arising due to coupling to the semi-infinite source/drain contacts and due to electron-phonon interaction, respectively. The imaginary part of the scattering self-energy function, $\text{Im}(\Sigma_{scat})$, is related to the in/out-scattering functions, $\Sigma_{scat}^{in/out}$, by, $\text{Im}(\Sigma_{scat}) = -i\left(\Sigma_{scat}^{in} + \Sigma_{scat}^{out}\right)/2$. The real part of $\Sigma_{scat}$ is estimated to be a minor correction, and is ignored in this work to minimize computational complexity [22].

The in/out-scattering functions for an optical phonon (OP) mode with energy $\hbar\omega$ are given by [27],

$$\Sigma_{scat}^{in,out}(E) = D_0\left(n_\omega + 1\right)G^{n,p}(E \pm \hbar\omega) + D_0 n_\omega G^{n,p}(E \mp \hbar\omega) \qquad (2)$$

where $D_0$ is the e-ph coupling parameter calculated according to [29], and $G^{n,p}$ are the electron/hole correlation functions given by [27], $G^{n,p} = G\Sigma^{in,out}G^\dagger$, where $\Sigma^{in,out}$ have contributions from both e-ph scattering (eq. (2)) as well as coupling to the contact reservoirs. Here we are assuming the scattering functions to be diagonal due to the local interaction approximation [22]. OP scattering by 190meV longitudinal optical (LO) mode, 180meV zone-boundary (ZBO) mode, and 21meV radial-breathing mode (RBM) have been considered. Acoustic phonon (AP) scattering is by the longitudinal acoustic (LA) mode. The relevant e-ph coupling parameters for the (16,0) CNT can be found in [22]. The phonon population is assumed to be in thermal equilibrium, with the number, $n_\omega$, given by the Bose-Einstein distribution,

$$n_\omega = \left(\exp(\hbar\omega/k_B T) - 1\right)^{-1} \qquad (3)$$

where $k_B$ is the Boltzmann constant and $T$ the device temperature. The first term in the right hand side of Eq. (2) corresponds to phonon emission mediated processes while the second term corresponds to phonon absorption. Finally, the current through the device from site $z$ to ($z+1$) in the nearest-neighbor tight-binding scheme is given by [22, 30],

$$I_{z \to z+1} = \frac{4ie}{\hbar} \int_{-\infty}^{+\infty} \frac{dE}{2\pi} \left[ H_{pz}(z,z+1)G^n(z+1,z) - H_{pz}(z+1,z)G^n(z,z+1) \right] \qquad (4)$$

where the lower and upper diagonal elements of the Hamiltonian and the electron correlation function have been used. The efficient numerical algorithms of [30] have been employed in our computational simulations.

The above-threshold (on-state) operation is discussed first. Figure 2 (a) compares the output characteristics, $I_{DS}$-$V_{DS}$, of the CNT p-i-n TFET under ballistic and dissipative transport. Here, it is seen that the influence of OP scattering (green dash-dot) becomes important only at large gate biases, even though the high-energy modes (LO and ZBO) have the strongest e-ph coupling [22]. Up to moderate gate biases, the current reduction is mainly due to carrier backscattering by AP and low-energy (RBM) phonon modes. Similar behavior on the influence of phonon scattering on the above-threshold operation of CNT-MOSFETs has already been reported [22, 31], and can be easily understood as follows. In order to have current reduction due to scattering by high-energy phonons, a majority of forward going carriers (source $\rightarrow$ drain) should be backscattered into empty backward going states by OP emission. Thus, the energy requirements for effective backscattering depends on the condition, $(E_{FS} - E_{C-channel}) = \eta_{FS} \geq \hbar\omega$, where $E_{FS}$ and $E_{C-channel}$ are the source Fermi level, and the channel conduction band position, respectively (see Figure 2 (b)) [31]. This is further exemplified in the energy-position resolved current density spectrum shown in Figure 2 (b). After reaching the drain region, however, carriers can be efficiently scattered by OP emission down to empty low-lying states, and thus will not possess enough energy to surmount the channel barrier and reach the source region again. On the other hand, elastic scattering due to acoustic and low-energy phonons can effectively backscatter at all gate biases, and is the dominant mechanism until high-energy OP scattering becomes effective at larger gate biases.

Therefore, in Figure 2(a) above-threshold performance is only moderately affected by phonon scattering. We observe, however, a lower value of the on-current (by about 4~10x) in the CNT p-i-n TFET compared to that of conventional CNT-MOSFETs [22, 31] due to the presence of the tunneling barrier. This observation confirms that the on-state performance of p-i-n TFETs is mainly dominated by the tunneling barrier properties, and moderately affected by the channel itself [5, 7, 19].

The subthreshold (off-state) operation is discussed next. Figure 3 (a) shows the transfer characteristics, $I_{DS}$-$V_{GS}$, for the CNT p-i-n TFET. Note that only the 180meV ZBO phonon mode, with strongest e-ph coupling, is included here since the subthreshold properties are observed to be mainly determined by this mode. In Figure 3 (a) it is seen that the p-i-n TFET has ambipolar behavior, which can be easily understood by observing Figure 2 (b). For small gate biases (or negative biases), the bands in the channel are pulled up, and a tunneling path appears at the channel-drain junction leading to ambipolar transport. As expected, the ambipolar branch appears earlier for larger drain biases. For identical source/drain doping concentrations the $I_{DS}$-$V_{GS}$ curves are symmetric around the minimum point. Asymmetric doping schemes can, however, suppress the ambipolar branch up to larger gate biases and lead to more desirable device characteristics [16]. It should be noted that, unlike in a conventional MOSFET, the subthreshold swing (*S*) of a p-i-n TFET is not a constant, but is bias dependent as seen in Figure 3(a) [5, 6, 19]. Interestingly, we observe less than 60mV/dec *S* at room temperature for both ballistic and dissipative transport (blue solid and red dashed curves); in our model device, *S* < 60mV/dec is obtained for $V_{GS}$ < 0.32V (left of dashed vertical line).

In Figure 3 (a) it is evident that the presence of OPs has a significant detrimental effect on the subthreshold properties; a minimum $S$ of 20mV/dec for ballistic transport (at $V_{DS}$ = 0.1V) degrades to 35mV/dec in the presence of OPs at room temperature ($T$ = 300K). The degradation is even greater at higher device temperatures; minimum $S \geq$ 40mV/dec at $T$ = 400K ($V_{DS}$ = 0.1V). The main reason for such deterioration of $S$ is due to phonon absorption assisted transport as exemplified in Figure 3 (b). Under the ballistic approximation, the minimum off-state current is mainly due to direct source to drain tunneling through the channel barrier region, and can be made very small; longer channel lengths, $L_{ch}$, further suppresses the off-current, and $S <$ 10mV/dec can be achieved for ballistic transport. Under such biasing conditions, however, higher-order processes such as phonon absorption assisted transport become significant, and tend to dominate the off-state characteristics. From Eq. (2) it is seen that the phonon absorption assisted process is proportional to the phonon occupation number, $n_\omega$, and increases at higher device temperatures (Eq. (3)). Thus, even though the direct tunneling processes are expected to be fairly temperature independent [5, 6, 19], higher-order processes such as phonon assisted transport that are temperature dependent can become important in the off-state of a BTBT device as seen here. They could also limit the desirable off-state characteristics that could have been achieved otherwise. At the same time, the on-state performance of the CNT p-i-n TFET is observed to be relatively temperature insensitive, and can be attributed to the fact that the above-threshold transport is mainly due to direct tunneling. In indirect bandgap semiconductors, such as technologically important silicon and germanium, BTBT would be mainly due to phonon assisted transport [32], and could have strong temperature dependence both in the off-state as well as the on-state.

In conclusion, we observe less than 60mV/dec subthreshold swings in p-i-n TFETs, in contrast to that in conventional (e.g. n-i-n) MOSFETs operating in the over-the-barrier conduction regime. The on-current of p-i-n TFETs is, however, mainly dependent on the tunneling barrier properties, and phonon scattering has only a moderate effect. On the other hand, the subthreshold operation is dominated by phonon assisted transport, and exhibits significant temperature dependence. Nevertheless, the low subthreshold swing is robust and persists even with the inclusion of phonon scattering and under higher source-to-drain biases.

*Acknowledgment* - S.O.K thanks the Intel Foundation for PhD Fellowship support. Computational support was provided by the NSF Network for Computational Nanotechnology (NCN).

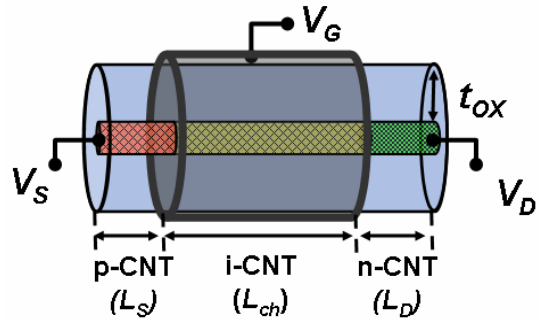

Figure 1. Modeled CNT p-i-n TFET structure with wrap-around gate used in this study. A (16,0) zigzag CNT with high-k $HfO_2$ ($k$ = 16, $t_{ox}$ = 2nm) gate oxide, intrinsic channel ($L_{ch}$ = 20nm), doped source (p-type, $N_S$ = 0.75/nm, $L_S$ = 35nm) and drain (n-type, $N_D$ = 0.75/nm, $L_D$ = 35nm) have been used. Source/drain doping levels can be compared to the carbon atom density in a (16,0) CNT of 150.2/nm.

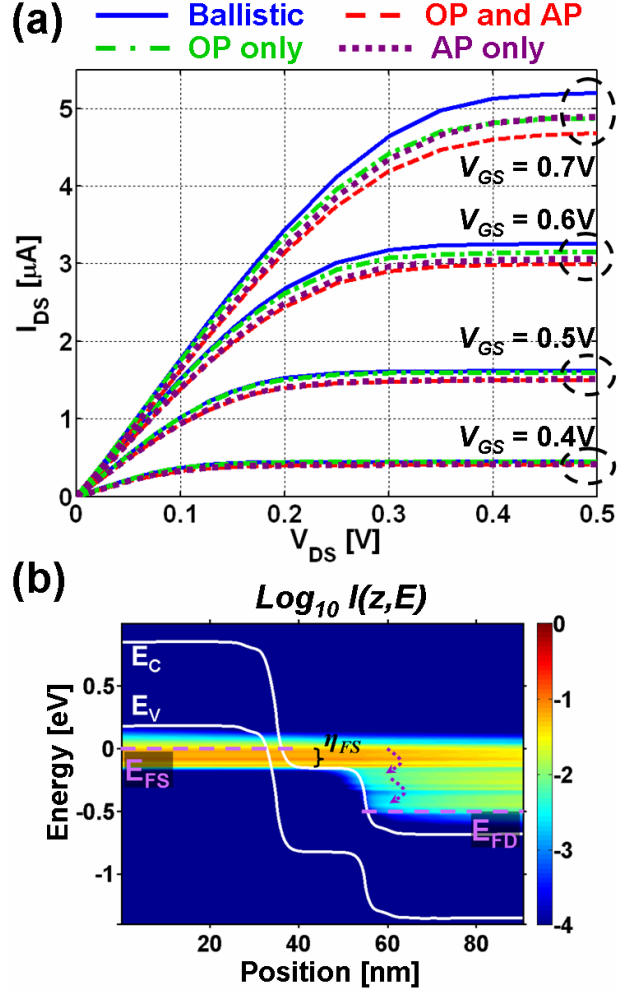

Figure 2. (a) Above-threshold $I_{DS}$-$V_{DS}$ characteristics for the CNT p-i-n TFET in Fig. 1. All the optical phonon modes (LO, ZBO, and RBM) are considered simultaneously for OP scattering. LA mode is considered for AP scattering. (b) Energy-position resolved current density spectrum (i.e., integrand of Eq. (4)) at $V_{DS} = V_{GS} = 0.5$V confirms the reduced influence of high-energy OP scattering on DC current transport up to moderate gate biases.

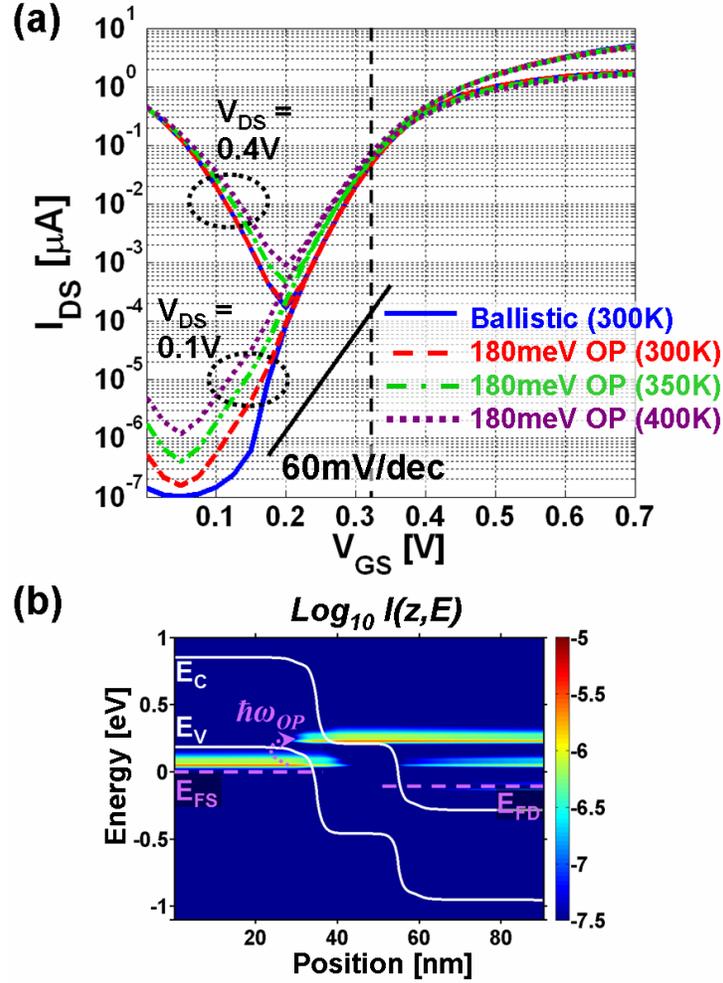

Figure 3. (a) $I_{DS}$-$V_{GS}$ characteristics for the CNT p-i-n TFET in Fig. 1 at $V_{DS}$ = 0.1V and 0.4V with and without phonon scattering. Simulated device temperatures are as shown in the legend. It is observed that the presence of phonons degrades the subthreshold swing; Higher device temperatures stronger degrade *S*, but sub-60mV/dec operation is consistently observed. (b) Energy-position resolved current density spectrum at a device temperature of 400K ($V_{GS}$ = 0.125V, $V_{DS}$ = 0.1V). Phonon absorption assisted transport is clearly observed at the source-channel junction. Carriers thermalize after reaching the drain.

References

[1]  *International Technology Roadmap for Semiconductors (ITRS)*: www.itrs.net.
[2]  S. Banerjee, W. Richardson, J. Coleman, and A. Chatterjee, "A new three-terminal tunnel device," *IEEE Electron Device Letters,* vol. ED-8, pp. 347-9, 1987.
[3]  T. Baba, "Proposal for surface tunnel transistors," *Japanese Journal of Applied Physics, Part 2: Letters,* vol. 31, pp. 455-457, 1992.
[4]  W. Hansch, C. Fink, J. Schulze, and I. Eisele, "Vertical MOS-gated Esaki tunneling transistor in silicon," *Thin Solid Films,* vol. 369, pp. 387-389, 2000.
[5]  K. K. Bhuwalka, J. Schulze, and I. Eisele, "Scaling the vertical tunnel FET with tunnel bandgap modulation and gate workfunction engineering," *IEEE Transactions on Electron Devices,* vol. 52, pp. 909-17, 2005.
[6]  Q. Zhang, W. Zhao, and A. Seabaugh, "Low-subthreshold-swing tunnel transistors," *IEEE Electron Device Letters,* vol. 27, pp. 297-300, 2006.
[7]  A. S. Verhulst, W. G. Vandenberghe, K. Maex, and G. Groeseneken, "Tunnel field-effect transistor without gate-drain overlap," *Applied Physics Letters,* vol. 91, p. 053102, 2007.
[8]  T. Uemura and T. Baba, "First demonstration of a planar-type surface tunnel transistor (STT): lateral interband tunnel device," *Solid-State Electronics:An International Journal,* vol. 40, pp. 519-522, 1996.
[9]  W. M. Reddick and G. A. Amaratunga, "Silicon surface tunnel transistor," *Applied Physics Letters,* vol. 67, p. 494, 1995.
[10] J. Koga and A. Toriumi, "Negative differential conductance at room temperature in three-terminal silicon surface junction tunneling device," *Applied Physics Letters,* vol. 70, p. 2138, 1997.
[11] C. Aydin, A. Zaslavsky, S. Luryi, S. Cristoloveanu, D. Mariolle, D. Fraboulet, and S. Deleonibus, "Lateral interband tunneling transistor in silicon-on-insulator," *Applied Physics Letters,* vol. 84, pp. 1780-1782, 2004.
[12] T. Nirschl, P. F. Wang, C. Weber, J. Sedlmeir, R. Heinrich, R. Kakoschke, K. Schrufer, J. Holz, C. Pacha, T. Schulz, M. Ostermayr, A. Olbrich, G. Georgakos, E. Ruderer, W. Hansch, and D. Schmitt-Landsiedel, "The Tunneling Field Effect Transistor (TFET) as an add-on for ultra-low-voltage analog and digital processes," *Technical Digest - International Electron Devices Meeting, IEDM,* pp. 195-198, 2004.
[13] K. R. Kim, H. H. Kim, K.-W. Song, J. I. Huh, J. D. Lee, and B.-G. Park, "Field-Induced Interband Tunneling Effect Transistor (FITET) with negative-differential transconductance and negative-differential conductance," *IEEE Transactions on Nanotechnology,* vol. 4, pp. 317-321, 2005.
[14] J. Appenzeller, Y. M. Lin, J. Knoch, and P. Avouris, "Band-to-band tunneling in carbon nanotube field-effect transistors," *Physical Review Letters,* vol. 93, p. 4, Nov 2004.

[15] J. Appenzeller, L. Yu-Ming, J. Knoch, C. Zhihong, and P. Avouris, "Comparing carbon nanotube transistors - the ideal choice: a novel tunneling device design," *IEEE Transactions on Electron Devices,* vol. 52, pp. 2568-76, 2005.

[16] S. O. Koswatta, D. E. Nikonov, and M. S. Lundstrom, "Computational study of carbon nanotube p-i-n tunnel FETs," *Technical Digest - International Electron Devices Meeting, IEDM,* pp. 518-521, 2005.

[17] Y. R. Lu, S. Bangsaruntip, X. R. Wang, L. Zhang, Y. Nishi, and H. J. Dai, "DNA functionalization of carbon nanotubes for ultrathin atomic layer deposition of high kappa dielectrics for nanotube transistors with 60 mV/decade switching," *Journal of the American Chemical Society,* vol. 128, pp. 3518-3519, Mar 2006.

[18] Z. Guangyu, W. Xinran, L. Xiaolin, L. Yuerui, J. Ali, and D. Hongjie, "Carbon nanotubes: from growth, placement and assembly control to 60mV/decade and sub-60 mV/decade tunnel transistors," *Technical Digest - International Electron Devices Meeting, IEDM,* p. 4 pp., 2006.

[19] J. Knoch, S. Mantl, and J. Appenzeller, "Impact of the dimensionality on the performance of tunneling FETs: Bulk versus one-dimensional devices," *Solid-State Electronics,* vol. 51, pp. 572-578, Apr 2007.

[20] R. Saito, G. Dresselhaus, and M. S. Dresselhaus, *Physical Property of Carbon Nanotubes*. London: Imperial College Press, 1998.

[21] J. Guo, "A quantum-mechanical treatment of phonon scattering in carbon nanotube transistors," *Journal of Applied Physics,* vol. 98, p. 6, Sep 2005.

[22] S. O. Koswatta, S. Hasan, M. S. Lundstrom, M. P. Anantram, and D. E. Nikonov, "Non-equilibrium Green's function treatment of phonon scattering in carbon nanotube transistors," *Ieee Transactions on Electron Devices,* vol. 54, pp. 2339-2351, 2007.

[23] A. Javey, J. Guo, D. B. Farmer, Q. Wang, E. Yenilmez, R. G. Gordon, M. Lundstrom, and H. J. Dai, "Self-aligned ballistic molecular transistors and electrically parallel nanotube arrays," *Nano Letters,* vol. 4, pp. 1319-1322, Jul 2004.

[24] S. O. Koswatta, M. S. Lundstrom, and D. E. Nikonov, "Band-to-Band Tunneling in a Carbon Nanotube Metal-Oxide-Semiconductor Field-Effect Transistor Is Dominated by Phonon-Assisted Tunneling," *Nano Lett.,* vol. 7, pp. 1160-1164, 2007.

[25] M. Pourfath, H. Kosina, and S. Selberherr, "Rigorous modeling of carbon nanotube transistors," *Journal of Physics: Conference Series,* vol. 38, pp. 29-32, 2006.

[26] S. O. Koswatta, M. S. Lundstrom, M. P. Anantram, and D. E. Nikonov, "Simulation of phonon-assisted band-to-band tunneling in carbon nanotube field-effect transistors," *Applied Physics Letters,* vol. 87, p. 3, Dec 2005.

[27] S. Datta, *Quantum Transport: Atom to Transistor*. Cambridge, U.K: Cambridge University Press, 2005.

[28] R. Venugopal, Z. Ren, S. Datta, M. S. Lundstrom, and D. Jovanovic, "Simulating quantum transport in nanoscale transistors: Real versus mode-space approaches," *Journal of Applied Physics,* vol. 92, p. 3730, 2002.

[29] S. Hasan, M. A. Alam, and M. S. Lundstrom, "Simulation of Carbon Nanotube FETs Including Hot-Phonon and Self-Heating Effects," *Electron Devices, IEEE Transactions on,* vol. 54, pp. 2352-2361, 2007.

[30] A. Svizhenko, M. P. Anantram, T. R. Govindan, B. Biegel, and R. Venugopal, "Two-dimensional quantum mechanical modeling of nanotransistors," *Journal of Applied Physics,* vol. 91, pp. 2343-54, 2002.

[31] S. O. Koswatta, S. Hasan, M. S. Lundstrom, M. P. Anantram, and D. E. Nikonov, "Ballisticity of nanotube field-effect transistors: Role of phonon energy and gate bias," *Applied Physics Letters,* vol. 89, p. 3, Jul 2006.

[32] S. Sedlmaier, J. Schulze, T. Sulima, C. Fink, C. Tolksdorf, A. Bayerstadler, I. Eisele, P. F. Wang, K. Hilsenbeck, and W. Hansch, "Phonon assisted tunneling in gated p-i-n diodes," *Materials Science and Engineering B: Solid-State Materials for Advanced Technology,* vol. 89, pp. 116-119, 2002.